\documentclass[preprint,aps]{revtex4}

\usepackage [cp1251]{inputenc}
\usepackage [english]{babel}
\makeatletter\AtBeginDocument{\let\@elt\relax}\makeatother
\usepackage{amsmath}
\usepackage{bm}
\usepackage{epsfig}
\usepackage{color}
\usepackage[unicode=true,colorlinks=true,citecolor=blue,urlcolor=blue]{hyperref}

\topmargin - 2.5 cm 

\usepackage{braket}
\graphicspath{{img/}}

\newcommand{\e}{\mathrm{e}}
\renewcommand{\d}{\mathrm d}

\begin{document}
\title{Interplay between hyperfine and anisotropic exchange interactions \\in exciton luminescence of quantum dots\footnote{This is a translation of the original manuscript in Russian in memory of V. M. Agranovich, which is available at \href{https://www.dropbox.com/scl/fi/kyyyaaimh2bjtf1dycv11/Agranovich_memorium.pdf?rlkey=ix0t7gi0ytxu5et4yhw134ke8&st=yp7n4f3o&dl=1}{this URL}.}}

\author{D. S. Smirnov}
\email{smirnov@mail.ioffe.ru}
\affiliation{Ioffe Institute, 194021 St. Petersburg, Russia}

\author{E. L. Ivchenko}
\affiliation{Ioffe Institute, 194021 St. Petersburg, Russia}

\begin{abstract}
  The optical orientation and alignment of excitons in semiconductor indirect gap quantum dots have been studied theoretically. A special regime is analyzed in which the energy of the hyperfine interaction of an electron with lattice nuclei is small compared to the exchange splitting between bright and dark excitonic levels, but is comparable to the anisotropic exchange splitting of the radiative doublet. The dependencies of degrees of circular and linear polarization on the external magnetic field under resonant excitation of excitons by polarized light are calculated.
\end{abstract}

\maketitle

\section{Introduction}

The polarization of excitonic luminescence excited in semiconductor quantum dots by polarized optical radiation is influenced by several parameters: exchange splitting $\delta_0$ between radiative and non-radiative doublets, anisotropic exchange splitting of each of these doublets $\delta_b$ and $\delta_d$, respectively, see e.g.~\cite{IvchenkoReview}, radiative and non-radiative lifetimes, $\tau_r$ and $\tau_{nr}$, and the energy of hyperfine interaction of an electron with the lattice nuclei, $\varepsilon_N$. In quantum dots grown on the basis of direct gap semiconductors, the energy $\varepsilon_N$ is small compared to $\delta_0$ and $\hbar/\tau_r$, and nuclear spins do not affect the photoluminescence of a bright exciton in magnetic fields at which the Zeeman splitting $\varepsilon_B$ of excitonic sublevels is small compared to $\delta_0$. In Ref.~\cite{SmirnovIvchenko} we studied polarized photoluminescence of indirect gap quantum dots in a special regime, where the exchange splitting $\delta_0$, the hyperfine interaction energy $\varepsilon_N$, and the radiative broadening $\hbar/\tau_r$ are comparable in order of magnitude, and therefore the Overhauser field created by nuclear spins plays an important role. In what follows, this regime is be designated by the Roman numeral I. 

There is another special regime of excitonic luminescence (regime II), in which (a) the exchange splitting is $\delta_0 \gg \varepsilon_N, \varepsilon_B$, but (b) the relationships between the quantities $\hbar/\tau_r$, $\varepsilon_N$, $ \delta_b$ and $\varepsilon_B$ are arbitrary. According to Ref.~\cite{Shamirzaev2019,DEP}, in an array of (In,Al)As/AlAs quantum dots, the distribution of their linear dimensions ensures the coexistence of regimes I and II for different dots in the same sample. The present work addresses the regime II of exciton luminescence, which has not been studied theoretically before.

The research relevance is related to the rapid development of the quantum technologies based on the phenomenon of entanglement. Quantum dots make it possible to generate and study entangled pairs of photons under cascade recombination of biexciton~\cite{XX1,XX2,XX3,XX4}, in particular, to experimentally test Bell's inequalities~\cite{entanglement-review}. The degree of two-photon entanglement is determined by the fine structure of the exciton~\cite{XX-review} and indeed can be limited by the hyperfine interaction in the case $\delta_b < \varepsilon_N$~\cite{XX-hf1,XX-hf2,XX-hf3}. However, as stated above, in conventional quantum dots, e.g. based on InGaAs/GaAs, the inequality $\hbar/\tau_r \gg \varepsilon_N$ holds and, in this case, the influence of nuclei is weak. We will show that in the regime II at $\varepsilon_B \ll \delta_0$, when the bright-dark exciton mixing can be neglected, the luminescence polarization can nevertheless be controlled by the hyperfine interaction. In the second part of the work, the case of stronger magnetic fields will be analyzed, in which the Zeeman splitting $\varepsilon_B$ exceeds $\delta_b, \varepsilon_N, \hbar/\tau_r$, and is comparable to $\delta_0$ giving rise to an anticrossing of sublevels of bright and dark excitons.

\section{Exciton Hamiltonian}

We consider a radiative doublet of an exciton composed of an electron in the X valley and a hole in the vicinity of the center $\Gamma$ of the Brillouin zone. The exciton is localized in an AlAs/AlGaAs quantum dot. The quantum dot boundary provides an admixture of the Bloch states near the $\Gamma$ point to the X electron wave function, which leads to an appearance of zero-phonon line of the exciton luminescence, in addition to the phonon-assisted recombination. For the large exchange splitting $\delta_0$ and in the absence of the exciton spin relaxation, the dark exciton states do not play any role. In the following, we use the laboratory coordinate frame $x_0,y_0,z$ with the $z$ axis along the structure growth axis. In addition, for each quantum dot we introduce the lateral coordinates $x,y$ in accordance with its anisotropy.

The two exciton sublevels $|+1 \rangle$ and $|-1 \rangle$, which are optically active in the $\sigma^+$ and $\sigma^-$ polarizations, can be conveniently described using the formalism of a pseudospin~\cite{Khitrova}. The Hamiltonian, which describes the fine structure of the exciton level, has the general form
\begin{equation} \label{OmegaS}
\mathcal H= \frac12 \hbar\bm\Omega \cdot \bm\sigma\:.
\end{equation}
Here $\bm\sigma$ is the vector composed of the Pauli matrices in this basis, and $\bm\Omega$ is an effective Larmor frequency of the pseudospin precession with the three components $\Omega_i$, $i = 1,2,3$.  The components $\Omega_1$ and $\Omega_2$ are determined by the anisotropy of the localizing potential taking into account the long-range exchange interaction~\cite{Gammon,Goupalov1996,Ivchenko1997,Goupalov1998,Stevenson,Voisin}. Obviously, the splitting $\delta_b$ defined in Introduction is equal to $\hbar \sqrt{\Omega^2_1 + \Omega_2^2}$. The third component is a sum
\begin{equation}
\Omega_3 = \Omega_N + \Omega_B\:,
\end{equation}
where the frequency $\Omega_N$ is related to fluctuations of the nuclear spins of crystal lattice and the frequency $\Omega_B$ equals $g_{\parallel} \mu_B B_z/\hbar$ with $g_{\parallel}$ being the exciton longitudinal $g$-factor, $\mu_B$ being the Bohr magneton, and $B_z$ being the $z$ component of the magnetic field ${\bm B}$. In the linear order, the lateral components of the magnetic field ${\bm B}$ and the Overhauser field do not enter the Hamiltonian~\eqref{OmegaS}. Here and in the next section, we consider the case of relatively weak fields $\hbar \Omega_B \ll \delta_0$.

The hyperfine interaction of electron and hole spins, ${\bm S}$ and ${\bm J}$, with the nuclei is short range and can be written in the form
\begin{equation}
  \mathcal H_{hf}= v_0 \sum_n \left[ \bm S \hat{A}_e \bm I_n \delta(\bm r_e-\bm R_n) + {\bm J} \hat{A}_h\bm I_n \delta(\bm r_h-\bm R_n)\right]\:,
\end{equation}
where the subscript $n$ enumerates nuclear spins $\bm I_n$ located at the nodes of the crystal lattice $\bm R_n$, $v_0$ is the volume of the elementary cell, $\bm r_{e,h}$ are the radius vectors of electron and hole within the smooth envelope approximation, $\hat{A}_{e,h}$ are the hyperfine interaction tensors which have the dimension of energy and, for simplicity, are assumed to be the same for all nuclei. Due to the time reversal symmetry, one can assume the envelopes of the electron and hole wave functions, $\Phi_e({\bm r}_e)$ and $\Phi_h({\bm r}_h)$, to be real. We assume that the size quantization energy of a single particle exceeds the exciton Rydberg. Then the nuclear field acting on the electron is given by the expression
\begin{equation}
  \hbar\Omega_N=v_0\sum_n \left[A_h \Phi_h^2(\bm R_n) - A_e \Phi_e^2(\bm R_n) \right]I_{n,z}\:,
\end{equation}
where $A_e=A_{e;zz}$, $A_h=3A_{h;zz}$ and the off diagonal components of the tensors $\hat{A}_e$ and $\hat{A}_h$ are set zero. Here we took into account that the exciton state $|\pm 1 \rangle$ is composed of a hole with the spin $\pm 3/2$ and an electron with the spin $\mp 1/2$.

We take into account that the nuclear spin dynamics takes place at the time scales much longer than the exciton lifetime and that the nuclear spins ${\bm I}_n$ are randomly oriented, because the effects of dynamic nuclear polarization are negligible. As a result, $\Omega_N$ is also random and  can be described by the Gaussian distribution function
\begin{equation} \label{distrib}
  \mathcal F(\Omega_N)=\sqrt{\frac{2}{\pi}}T_2^*\e^{-2(\Omega_NT_2^*)^2}\:,
\end{equation}
where the inverse dephasing time~\cite{Glazov}
\begin{equation} \label{T2*}
 \frac{1}{T_2^*} = \frac{v_0}{\hbar}\sqrt{\sum_n \frac{4}{3}I_n(I_n+1) \left[ A_h \Phi_h^2(\bm R_n) - A_e \Phi_e^2(\bm R_n) \right]^2}
\end{equation}
determines the dispersion of the random field: $\braket{\Omega_N^2}=1/(2T_2^*)^2$.

\section{Polarized luminescence}

We note that both the inverse radiative exciton lifetime $\tau^{-1}_r$ and the splitting of the radiative doublet $\delta_b=\hbar \Omega_\perp = \hbar \sqrt{\Omega_{x}^2+\Omega_{y}^2}$ determined by the long range exchange interaction are proportional to the squared matrix element of the exciton excitation and, in particular, to the squared overlap of the electron and hole wave functions in the real and momentum spaces~\cite{bookIvchenko}. Therefore, the overlap integral is cancelled in the product $\Omega_{\perp} \tau_r \equiv w_{QD}$. For a direct band quantum dot this product can be very large, $w_{QD} \gg 1$, or very small, $w_{QD} \ll 1$, depending on the anisotropy of its shape. So for the indirect band quantum dots, the possibility of such a spread in the values of $w_{QD}$ is still possible. Here we focus on the most interesting case $w_{QD} \gg 1$ neglecting nonradiative recombination and the spin relaxation unrelated to the hyperfine interaction~\cite{DEP}.

Let us introduce the vector $\bm P^{(0)}$ composed of the three Stokes parameters $P^{(0)}_1, P^{(0)}_2, P^{(0)}_3 $~\cite{LL2} of the light incident along the normal $z$ on the sample surface and a similar vector $\bm P$ for the light emitted forward by the excitons. Instead of the indices $l, l',c$~\cite{SmirnovIvchenko} we use the indices 1, 2, 3. For the emission in the backward geometry, one has to change the sign of the component $P_3$, which characterizes the circular polarization of light. At $w_{QD} \gg 1$, the sets of Stokes parameters are related by the equation
\begin{equation}
P_i = \sum_{j=1,2,3} \Lambda_{ij} P^{(0)}_j\:,
\end{equation}
with the coupling matrix~\cite{SmirnovIvchenko}
\begin{equation}
\Lambda_{ij} = \frac{\Omega_i \Omega_j}{\Omega^2}
\end{equation}
and $\Omega^2= \Omega_1^2 + \Omega_2^2 + \Omega_3^2$. Since the total photoluminescence intensity does not depend on the polarization of the exciting light, this expression can be averaged over the distribution~\eqref{distrib} of the nuclear field $\Omega_N$. For each of the quantum dots, one can choose the lateral axes $x,y$, such as $\Omega_2 = 0$ and $\Omega_{\perp}= \Omega_1 > 0$. Then the nonzero components of the coupling matrix of the Stokes parameters have the form
\begin{subequations}
  \label{eq:L}
  \begin{equation} \label{L11}
    \braket{\Lambda_{11}}= 1 - \braket{\Lambda_{33}} =\int\limits_{-\infty}^\infty\frac{\Omega_1^2 F(\Omega_N) \d\Omega_N}{\Omega_1^2+(\Omega_B+\Omega_N)^2} =\pi\Omega_1T_2^*V(\Omega_BT_2^*;1/2,\Omega_1 T_2^*),
  \end{equation}
  \begin{equation} \label{L13}
    \braket{\Lambda_{13}}=\braket{\Lambda_{31}}=\frac{\Omega_B}{\Omega_1}\braket{\Lambda_{33}},
  \end{equation}
 \end{subequations}
where the Voigt distribution (or profile) is defined as
\[
V(u;\sigma,\gamma) = \int\limits_{- \infty}^{\infty} \frac{{\rm e}^{- v^2/(2 \sigma^2)}}{\sigma \sqrt{2 \pi}} \frac{1}{\pi} \frac{\gamma}{(u - v)^2 + \gamma^2} dv\:.
\]
Thus, the polarization properties of the photoluminescence of a single quantum dot are described by the two dimensionless parameters $\Omega_1 T_2^*$ and $\Omega_B T_2^*$.

\begin{figure}
  \includegraphics[width=0.47\linewidth]{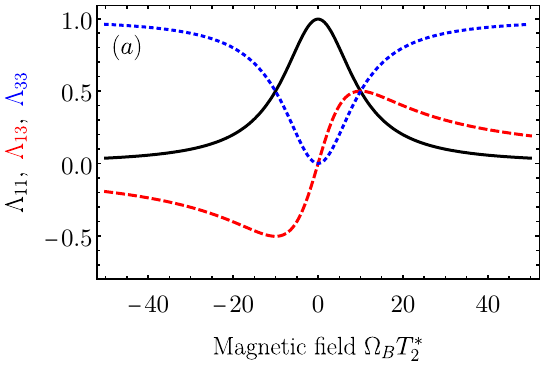}
  \hfill
  \includegraphics[width=0.47\linewidth]{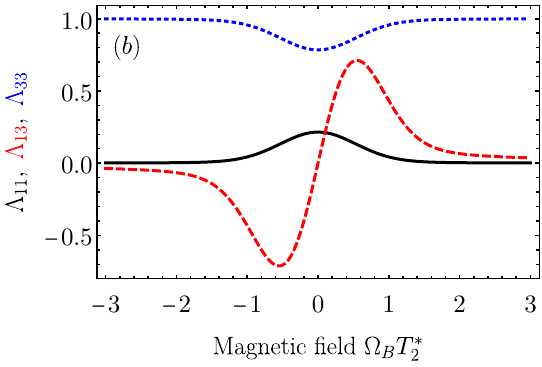}
  \caption{\label{fig} Coefficients of linear coupling between the Stokes parameters of the exciting light and the luminescence $\Lambda_{11}$ (black solid curve), $\Lambda_{13}$ (red dashed curve), $\Lambda_{33}$ (blue dotted curve) calculated after Eqs.~\eqref{eq:L} for $\Omega_1 T_2^*=10$ (a) and $0.1$ (b).}
\end{figure}

Magnetic field dependencies of the three nonzero components of the matrix $\braket{\hat{\Lambda}}$ are presented in Fig.~\ref{fig} for the cases of (a) strong, $\Omega_1 T_2^*\gg 1$, and (b) weak, $\Omega_1 T_2^*\ll 1$, anisotropic splitting. In the case (a), the hyperfine interaction does not play a significant role and the magnetic field dependencies hardly differ from those in the limit of $\Omega_N \to 0$~\cite{bookIvchenko}:
\[
\braket{\Lambda_{11}(\Omega_N \to 0)} = \frac{\Omega_1^2}{\Omega_{1}^2+ \Omega_B^2}\:,\:
\]
while the components $\braket{\Lambda_{33}(\Omega_N \to 0)}$ and $\braket{\Lambda_{13}(\Omega_N \to 0)}$ are expressed via $\braket{\Lambda_{11}(\Omega_N \to 0)}$ with the help of the relations~\eqref{L13}. In the case (b) the main role is played by the hyperfine interaction and Eq.~\eqref{L11} reduces to
\[
\braket{\Lambda_{11}(\Omega_1 \to 0)} = \pi \Omega_1 F(\Omega_B)\:.
\]
From comparison with the Fig.~\ref{fig}(a), one can see that the Overhauser field suppresses the optical alignment and enhances the optical orientation.

Let the axes $x,y$ be rotated counter clockwise by the angle $\phi$ relative to the laboratory axes $x_0, y_0$. Then the coupling matrix $\langle \Lambda_{i_0 j_0} \rangle$ in the frame $(x_0, y_0)$ is related to the components of the matrix $\langle \Lambda_{i j} \rangle$ by
\begin{equation}
\left\vert \left\vert \langle \Lambda_{i_0 j_0} \rangle \right\vert \right\vert = \left[ \begin{array}{ccc} \langle \Lambda_{11} \rangle \cos^2{2 \phi}   &  \langle  \Lambda_{11} \rangle \sin{2\phi} \cos{2\phi} &  \langle \Lambda_{13} \rangle \cos{2 \phi} \\  \langle  \Lambda_{11} \rangle \sin{2\phi} \cos{2\phi} & \langle  \Lambda_{11} \rangle \sin^2{2\phi} & \langle \Lambda_{13} \rangle \sin{2 \phi}\\  \langle \Lambda_{13} \rangle \cos{2 \phi}&  \langle \Lambda_{13} \rangle \sin{2 \phi} &  \langle \Lambda_{33} \rangle \end{array} \right]\:.
\end{equation}

If the lateral anisotropy axes of all the quantum dots in the ensemble coincide, one can use Eqs.~\eqref{L11}, \eqref{L13}, which should be averaged over the distribution of the splitting $\Omega_1$. If the lateral anisotropy axes are randomly distributed over all directions, then after averaging over the angle $\phi$, only the diagonal components of the coupling matrix remain:
\[
\braket{\braket{\Lambda_{11}}} = \braket{\braket{\Lambda_{22}}} = \frac12 \braket{\Lambda_{11}}\:,\:\braket{\braket{\Lambda_{33}}} = \braket{\Lambda_{33}}\:.
\]

\section{Anticrossing of exciton sublevels with increasing magnetic field strength}

Now we turn to a strong magnetic field, such that the energies of one of the dark and one of the bright excitons become equal, for example, for the exciton with the angular momentum projection $+1$~\cite{IvchenkoReview}. In this case, the Overhauser field is a small perturbation which not only lowers the symmetry of the system but also leads to a strong resonant mixing of the crossing levels, i.e.~to their anticrossing. This significantly affects the intensity and polarization of the luminescence even for the direct band quantum dots~\cite{Masumoto,BrightDark,mixing6}, see also Refs.~\onlinecite{mixing4,Boldyrev}. The effect of anticrossing due to the field $B_N$ can be observed at the magnetic fields $B^{({\rm cr})}_e = \delta_0/(2 \mu_B |g_{e,\parallel}|)$ and $B^{({\rm cr})}_h =  \delta_0/(2 \mu_B |g_{h,\parallel}|)$~\cite{IvchenkoReview}. Consider a single quantum dot in the field $B^{({\rm cr})}_e$ and demonstrate that due to the nuclear field under the resonant unpolarized excitation, the exciton emission from such a quantum dot gets partially circularly polarized and changes its intensity compared to that in the nonresonant field $B \neq B^{({\rm cr})}_e$.

At $B = B^{({\rm cr})}_e$ the sublevels with the angular momentum projections $+2$ and $+1$ can be considered as a two level system (we account for $g_{e,\parallel}>0$) with the Hamiltonian
\[
{\cal H} = \hbar {\bm \Omega}_N^e {\bm S}_e \:,
\]
where
\[
{\bm \Omega}_N^e = \frac{v_0}{\hbar} A_e \sum_n \Phi_e^2 ({\bm R}_n) {\bm I}_n\:,
\]
and we assume $A_{e;xx}=A_{e;yy}=A_e$.
 
Let us introduce the angle $\vartheta$ defined by
\[
\cos{\vartheta} = \frac{ \Omega_{N,z}^e }{ | {\bm \Omega}_{N}^e | } \:.
\]
Under the resonant excitation by unpolarized light, the intensity $I_{-1}$ of the exciton $\left|-1\right\rangle$ emission is proportional to the combination of times $\tau_{nr}/[\tau_r(\tau_r + \tau_{nr})]$ and does not change in the vicinity of $B^{({\rm cr})}_e$, while the intensity of exciton emission from the levels experiencing the anticrossing does depend on $\vartheta$. Skipping the details of the calculation, we present the ultimate result for the intensity of the exciton emission $I$ and the degree of circular polarization $P_{c}$ at the resonant unpolarized excitation
\begin{eqnarray} \label{IPc}
&& I \propto \frac{\tau_{nr}}{\tau_r(\tau_r + \tau_{nr})}\frac{a + b \cos^2{\vartheta}}{c -d \cos^2{\vartheta}}\:,\\ && P_c = \frac{\tau_r (2 \tau_r + \tau_{nr})}{a + b \cos^2{\vartheta}} \sin^2{\vartheta}\:.\label{abcd}
\end{eqnarray}
Here we introduced the coefficients
\begin{eqnarray} 
&& a = (2 \tau_r + \tau_{nr}) (3 \tau_r + 2 \tau_{nr})\:,\:b = 2 \tau^2_r + \tau_r \tau_{nr} - 2\tau^2_{nr}\:, \nonumber \\ && c = (2 \tau_r + \tau_{nr})^2\:,\:d = \tau^2_{nr}, \nonumber
\end{eqnarray}
which satisfy the relation $a + b = 2(c -d) = 8 \tau_r (\tau_r + \tau_{nr})$.

In analogy with Eq.~\eqref{T2*}, we define the relaxation time $T_{2,e}^*$ as follows
\begin{equation}
\frac{1}{T_{2,e}^*} = \frac{v_0}{\hbar} \sqrt{\sum_n \frac{4}{3}I_n(I_n+1)A_e^2\Phi_e^4(\bm R_n)}.
\end{equation}
In realistic structures, $\tau_{nr} \gg \tau_r$~\cite{DEP_details}. As a consequence, the dimensionless parameter
$$
\xi = \frac{\tau_r}{\tau_{nr}} \left( \delta_0 \frac{T_{2,e}^*}{\hbar} \right)^2
$$ 
can be both greater and smaller than unity. Equations~\eqref{IPc}, \eqref{abcd} are valid at $\xi \gg 1$~\cite{DEP_CISP}, while at $\xi \ll 1$, the dynamic electron spin polarization is realized~\cite{DEP}. Note that in those articles, analytical expressions are derived for the averaged intensity and circular polarization degree for the both limiting cases.

\section{Conclusion}

In the epitaxial semiconductor quantum dots, in addition to the ``bright-dark'' exciton exchange splitting $\delta_0$, there is a smaller exchange splitting of the bright exciton level $\delta_b$ caused by the local lateral anisotropy of the structure. The suppression of the long-range exchange interaction of electron and hole in the indirect quantum dots leads to an increasing role in the exciton fine structure of the hyperfine interaction with the nuclei of the crystal lattice. In this work, we have studied a special case of the exciton fine structure where the energy of the hyperfine interaction $\varepsilon_N$ is small compared to $\delta_0$ but the relation between $\varepsilon_N$ and $\delta_b$ is arbitrary. It is shown that in this special case, the nuclear spin fluctuations along the growth axis of the quantum dots structure lead to the suppression of the optical alignment effect and enhancement of the optical orientation. Dependencies of these effects on the longitudinal magnetic field are generally described by the Voigt profile. Also, we have analyzed the effect of the Overhauser field on the intensity and polarization of emission in the vicinity of the magnetic-induced anticrossing of the sublevels of bright and dark excitons.

\section*{Acknowledgments}

We acknowledge useful discussions with A.V. Rodina. The work is supported by the Russian Science Foundation grant N\textsuperscript{\underline{o}} 23-12-00142. D.S.S. is grateful to the Theoretical Physics and Mathematics Advancement Foundation ``BASIS''.



\begin{thebibliography}{50}
\bibitem{IvchenkoReview} E.L. Ivchenko, Magnetic circular polarization of exciton photoluminescence, Phys. Solid State {\bf 60}, 1514  (2018) [Fizika Tverd. Tela {\bf 60}, 1471 (2018)].
\bibitem{SmirnovIvchenko} D.V. Smirnov, E.L. Ivchenko, Theory of polarized luminescence of indirect band gap excitons in type-I quantum dots, Phys. Rev. B {\bf 108}, 195432 (2023).
\bibitem{Shamirzaev2019} J. Rautert, T.S. Shamirzaev, S.V. Nekrasov, D.R. Yakovlev, P. Klenovsk\'y, Yu.G. Kusrayev, M. Bayer, Optical orientation and alignment of excitons in direct and indirect band gap (In,Al)As/AlAs quantum dots with type-I band alignment, Phys. Rev. B \textbf{99}, 195411 (2019).
\bibitem{DEP} D.S. Smirnov, T.S. Shamirzaev, D.R. Yakovlev, M. Bayer, Dynamic polarization of electron spins interacting with nuclei in semiconductor nanostructures, Phys. Rev. Lett. \textbf{125}, 156801 (2020).
\bibitem{XX1} N. Akopian, N.H. Lindner, E. Poem, Y. Berlatzky, J. Avron, D. Gershoni, B.D. Gerardot, and P.M. Petroff, Entangled photon pairs from semiconductor quantum dots, Phys. Rev. Lett. {\bf 96}, 130501 (2006).
\bibitem{XX2} A. Dousse, J. Suffczynski, A. Beveratos, O. Krebs, A. Lemaitre, I. Sagnes, J. Bloch, P. Voisin, P. Senellart, Ultrabright source of entangled photon pairs, Nature {\bf 466}, 217 (2010).
\bibitem{XX3} D. Huber, M. Reindl, S.F. Covre da Silva, C. Schimpf, J. Martin-Sanchez, H. Huang, G. Piredda, J. Edlinger, A. Rastelli, R. Trotta, Strain-tunable GaAs quantum dot: A Nearly dephasing-free source of entangled photon pairs on demand, Phys. Rev. Lett. {\bf 121}, 033902
(2018).
\bibitem{XX4} F.T. {\O}stfeldt, E.M. Gonz\'alez-Ruiz, N. Hauff, Y. Wang, A.D. Wieck, A. Ludwig, R. Schott, L. Midolo, A.S. S{\o}rensen, R. Uppu, P. Lodahl, On-demand source of dual-rail photon pairs based on chiral interaction in a nanophotonic waveguide, PRX Quantum {\bf 3}, 020363 (2022).
\bibitem{entanglement-review} O. G{\"u}hne, G. T{\'o}th, Entanglement detection. Phys. Rep. {\bf 474}, 1 (2009).
\bibitem{XX-review} C. Schimpf, M. Reindl, B.B. Francesco, K.D. J{\"o}ns, R. Trotta, A. Rastelli, Quantum dots as potential sources of strongly entangled photons: Perspectives and challenges for applications in quantum networks, Appl. Phys. Lett. {\bf 118}, 100502 (2021).
\bibitem{XX-hf1} T. Kuroda, T. Mano, N. Ha, H. Nakajima, H. Kumano, B. Urbaszek, M. Jo, M. Abbarchi, Y. Sakuma, K. Sakoda, I. Suemune, X. Marie, T. Amand, Symmetric quantum dots as efficient sources of highly entangled photons: Violation of Bell's inequality without spectral and temporal filtering, Phys. Rev. B {\bf 88}, 041306(R) (2013).
\bibitem{XX-hf2} D. Huber, M. Reindl, Y. Huo, H. Huang, J.S. Wildmann, O.G. Schmidt, A. Rastelli, R. Trotta, Highly indistinguishable and strongly entangled photons from symmetric GaAs quantum dots, Nat. Commun. {\bf 8}, 15506 (2017).
\bibitem{XX-hf3} C. Schimpf, F.B. Basset, M. Aigner, W. Attenender, L. Gin\'es, G. Undeutsch, M. Reindl, D. Huber, D. Gangloff, E.A. Chekhovich, C. Schneider, S. H\"ofling, A. Predojevi\ifmmode \acute{c}\else \'{c}\fi{}, R. Trotta, A. Rastelli, Hyperfine interaction limits polarization entanglement of photons from semiconductor quantum dots, Phys. Rev. B {\bf 108}, L081405 (2023).
\bibitem{Khitrova} R.I. Dzhioev, H.M. Gibbs, E.L. Ivchenko, G. Khitrova, V.L. Korenev, M.N. Tkachuk, B.P. Zakharchenya, Determination of interface preference by observation of linear-to-circular polarization conversion under optical orientation of excitons in type-II GaAs/AlAs superlattices, Phys. Rev. B {\bf 56}, 13 405 (1997).
\bibitem{Goupalov1996} S.V. Goupalov, E.L. Ivchenko, A.V. Kavokin, Fine structure of excitonic levels in small anisotropic quantum systems, Proc. Int. Symp. ``Nanostructures: Physics and Technology'', St. Petersburg, 1996, pp. 322-325; Anisotropic exchange splitting of excitonic levels in small quantum systems, Superlatt. Microstruct. {\bf 23}, 1205 (1998).
\bibitem{Gammon}D. Gammon, E.S. Snow, B.V. Shanabrook, D.S. Katzer, and D. Park, Fine Structure Splitting in the Optical Spectra of Single GaAs Quantum Dots, Phys. Rev. Lett. {\bf 76}, 3005 (1996).
\bibitem{Ivchenko1997} E.L. Ivchenko, Fine structure of excitonic levels in semiconductor nanostructures, Phys. Status Solidi A {\bf 164}, 487 (1997).
\bibitem{Goupalov1998}    S.V. Gupalov, E L. Ivchenko and A.V. Kavokin, Fine structure of localized exciton levels in quantum wells, JETP {\bf 86}, 388 (1998) [Zh. Eksp. Teor. Fiz. {\bf 113}, 703 (1998)].
\bibitem{Stevenson} R.M. Stevenson, R.J. Young, P. See, D.G. Gevaux, K. Cooper, P. Atkinson, I. Farrer, D.A. Ritchie, A.J. Shields, Magnetic-field-induced reduction of the exciton polarization splitting in InAs quantum dots, Phys. Rev. B {\bf 73}, 033306 (2006).
\bibitem{Voisin} M.M. Glazov, E.L. Ivchenko, O. Krebs, K. Kowalik, P. Voisin, Diamagnetic contribution to the effect of in-plane magnetic field on a quantum-dot exciton fine structure, Phys. Rev. B {\bf 76},
193313 (2007).
\bibitem{Glazov} M.M. Glazov, Electron \& Nuclear Spin Dynamics in Semiconductor
Nanostructures (Oxford University Press, Oxford, 2018).
\bibitem{bookIvchenko} E.L.~Ivchenko, Optical Spectroscopy of Semiconductor Nanostructures (Alpha Science International, Harrow, UK, 2005).
\bibitem{LL2} L.D. Landau, E.M. Lifshitz, Course of Theoretical Physics, Vol. 2, The Classical Theory of Fields, Butterworth Heinemann, 4th Edition, 1994.
\bibitem{Masumoto}Y. Masumoto, K. Toshiyuki, T. Suzuki, M. Ikezawa, Resonant spin orientation at the exciton level anticrossing in InP quantum dots, Phys. Rev. B {\bf 77}, 115331 (2008).
\bibitem{BrightDark} H. Kurtze, D.R. Yakovlev, D. Reuter, A.D. Wieck, M. Bayer, Hyperfine interaction mediated exciton spin relaxation in (In,Ga)As quantum dots, Phys. Rev. B {\bf 85}, 195303 (2012).
\bibitem{mixing6} D. Cogan, O. Kenneth, N.H. Lindner, G. Peniakov, C. Hopfmann, D. Dalacu, P.J. Poole, P. Hawrylak, D. Gershoni, Depolarization of Electronic Spin Qubits Confined in Semiconductor Quantum Dots, Phys. Rev. X {\bf 8}, 041050 (2018).
\bibitem{mixing4} A.N. Starukhin, D.K. Nelson, B.S. Razbirin, D.L. Fedotov, D.K. Syunyaev, Evolution of the level anticrossing signal in magnetoluminescence of localized excitons in the GaSe-GaTe solid solution, Phys. Solid State {\bf 57}, 1937 (2015)  [Fizka Tverd. Tela {\bf 57}, 1888 (2015)].
\bibitem{Boldyrev} K.N. Boldyrev, M.N. Popova, B.Z. Malkin, N.M. Abishev, Direct observation of hyperfine level anticrossings in the optical spectra of a $^7$LiYF$_4$:Ho$^{3+}$ single crystal, Phys. Rev. B {\bf 99}, 041105 (2019).
\bibitem{DEP_details} T.S. Shamirzaev, A.V. Shumilin, D.S. Smirnov, J. Rautert, D.R. Yakovlev, M. Bayer, Dynamic polarization of electron spins in indirect band gap (In,Al)As/AlAs quantum dots in a weak magnetic field: Experiment and theory, Phys. Rev. B {\bf 104}, 115405 (2021).
\bibitem{DEP_CISP} A.V. Shumilin, T.S. Shamirzaev, D.S. Smirnov, Spin light emitting diode based on exciton fine structure tuning in quantum dots, Phys. Rev. Lett. {\bf 132}, 076202 (2024).
\end{thebibliography}
\end{document}